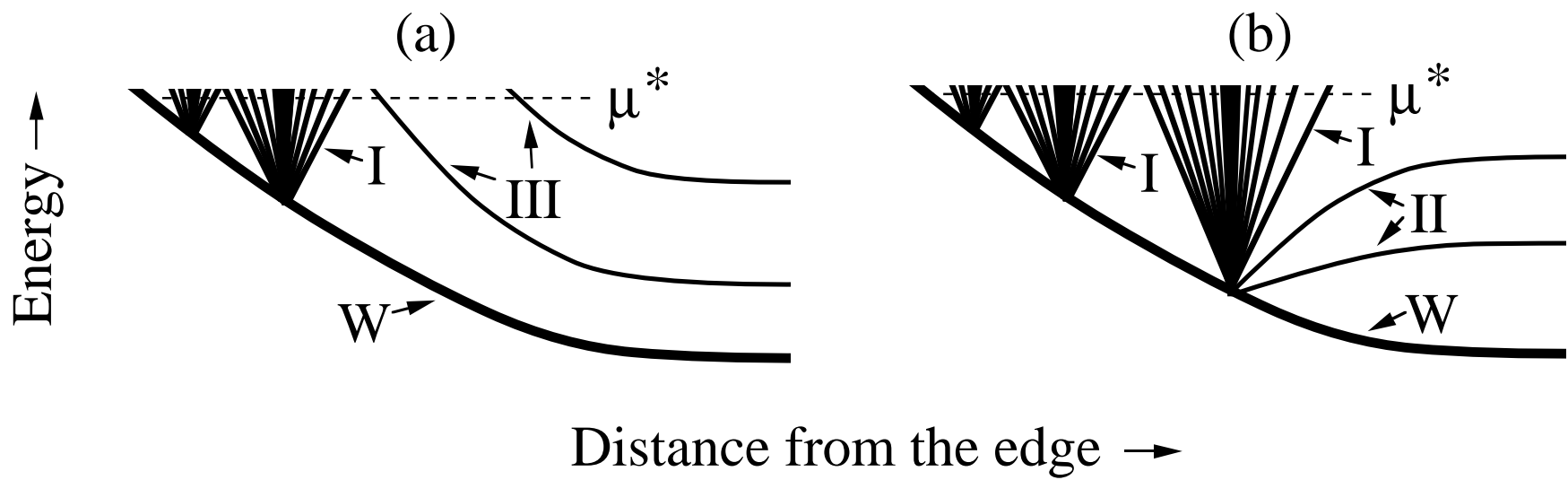

Figure 1
Kirczenow

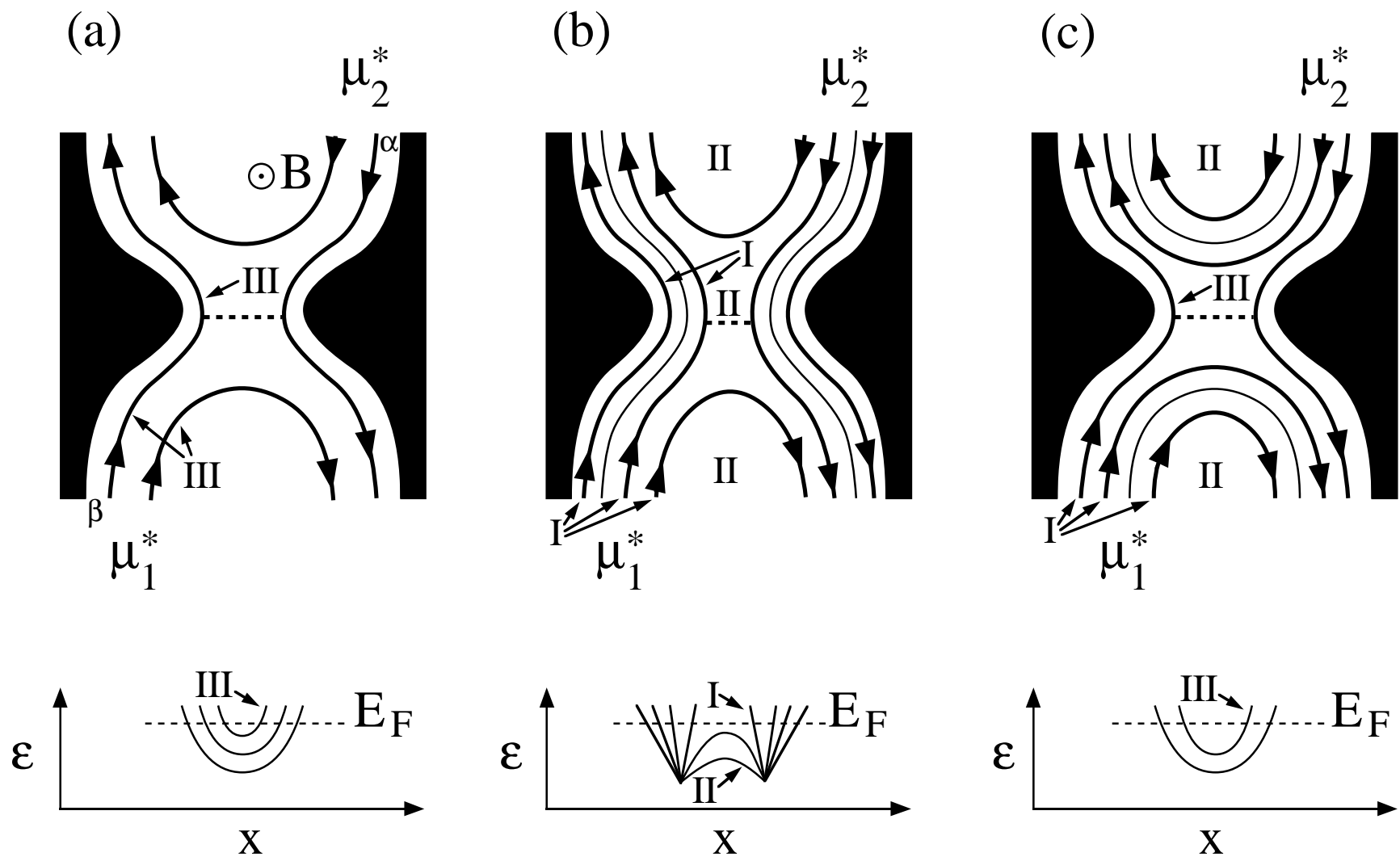

Figure 2
Kirczenow

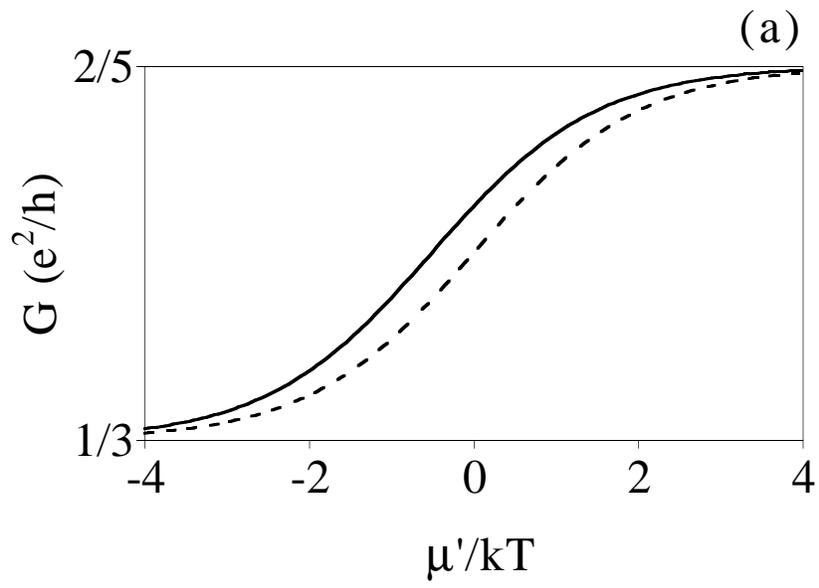

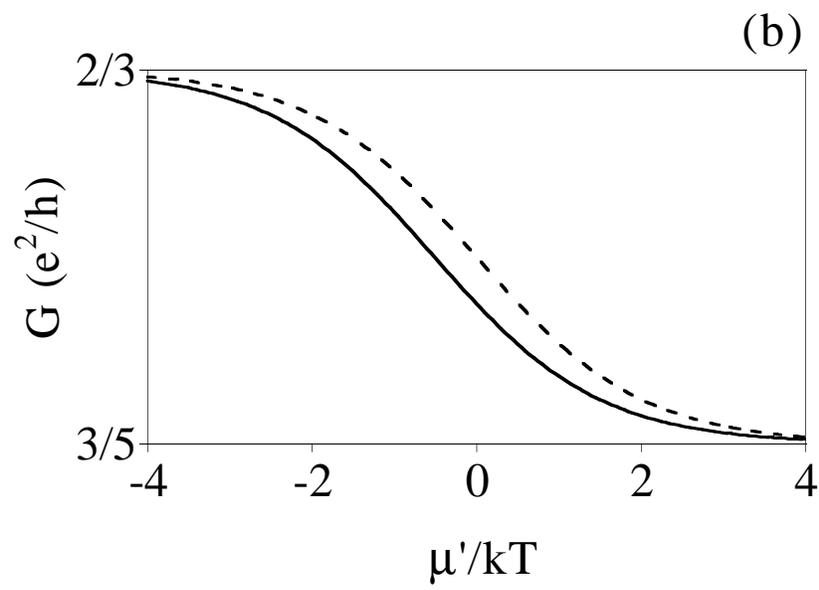

Figure 3
Kirczenow

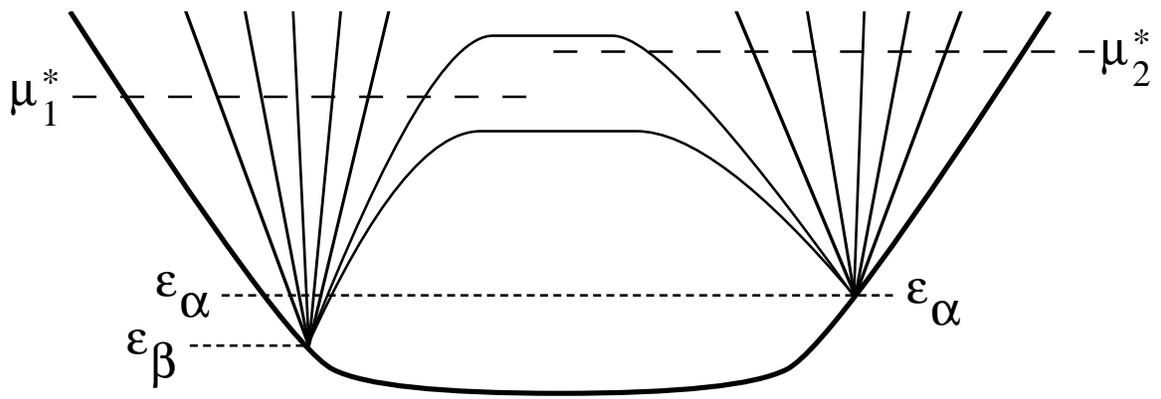

Figure 4
Kirczenow

Figure 5
Kirczenow

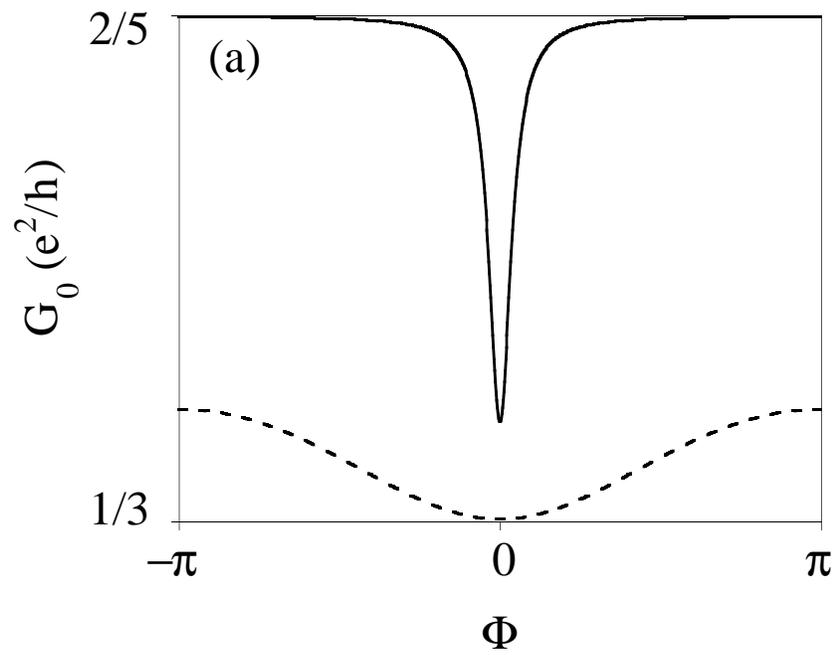

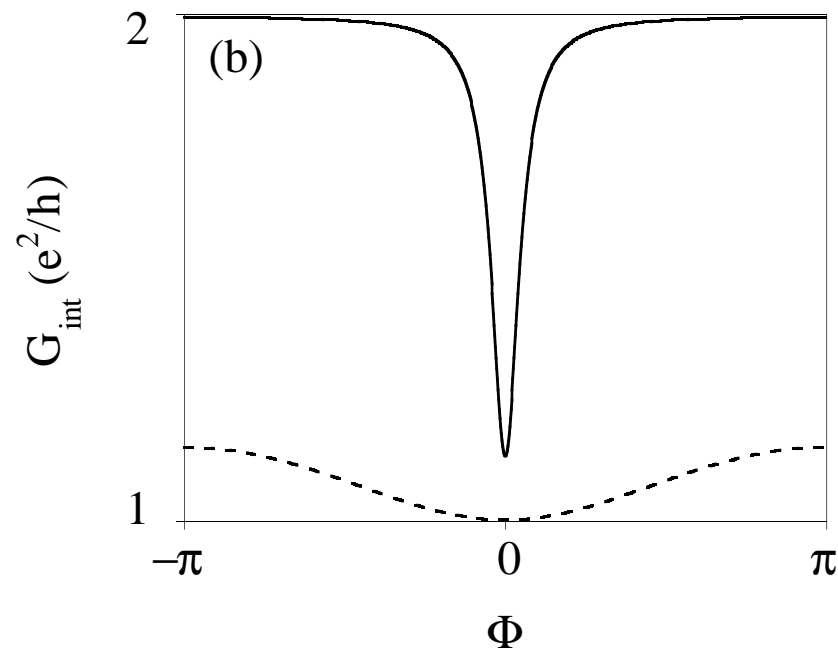

Figure 6
Kirczenow

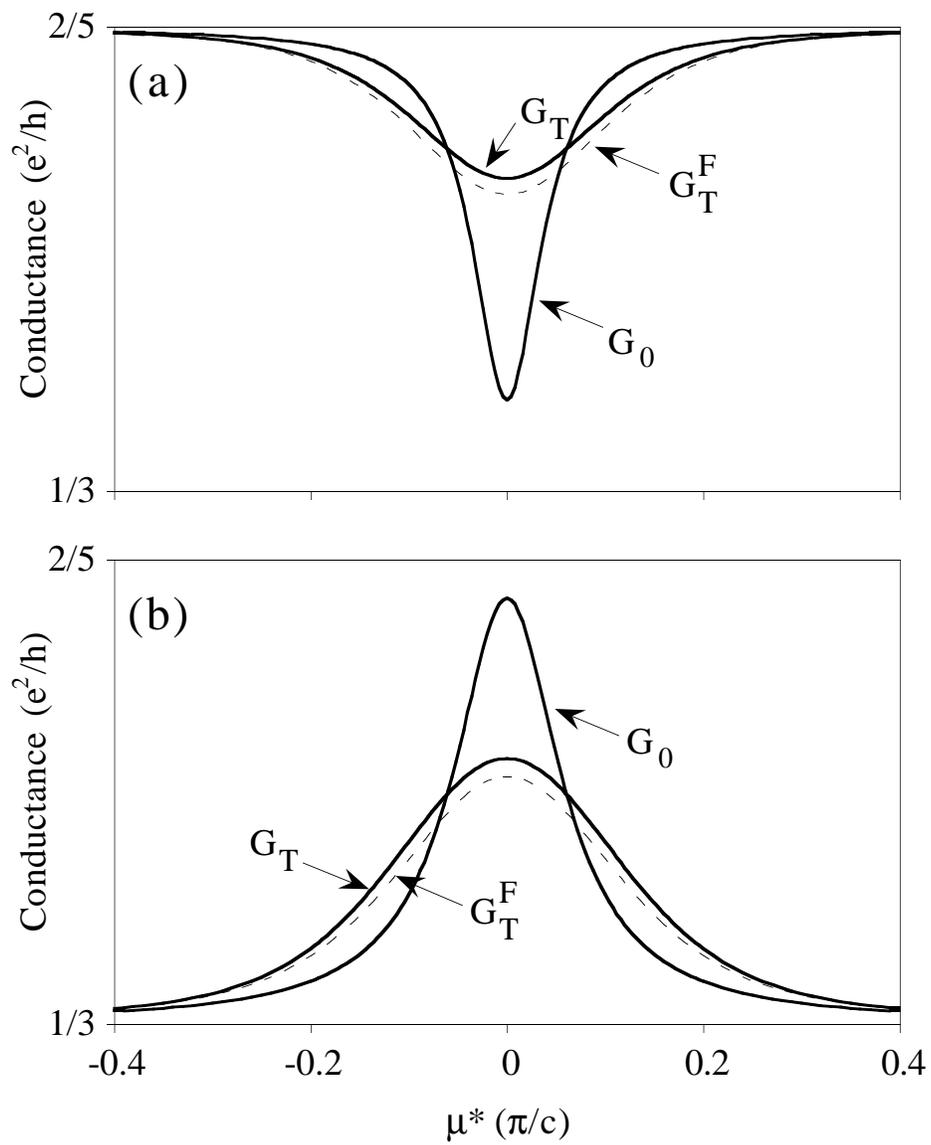

Figure 7
Kirczenow

# Composite Fermion Edge States and Transport Through Nanostructures in the Fractional Quantum Hall Regime

George Kirczenow

*Department of Physics, Simon Fraser University,
Burnaby, British Columbia,
Canada, V5A 1S6*

A theory of transport through semiconductor nanostructures in the fractional quantum Hall regime is proposed, based on a model of composite fermion edge states. Adiabatic and non-adiabatic constrictions and constrictions containing artificial impurities are studied as examples. The results obtained, including the temperature-dependent behavior of Aharonov-Bohm resonances in the fractional regime, are in good agreement with experiments. The temperature dependence predicted by the composite fermion theory for features in the two-terminal conductance of both adiabatic constrictions and constrictions with artificial impurities, is close to that expected from ordinary Fermi liquid phenomenology. However there are significant differences that should be detectable by careful measurements. Some differences between the present results and those obtained from Luttinger liquid models are discussed in the context of the available experimental data.

## 1. Introduction

It was first shown by Laughlin,[1] that the fractional quantum Hall effect occurs because electron-electron interactions result in incompressible states of two-dimensional electron systems at special filling fractions of a Landau level. Subsequently, Jain showed[2] that these incompressible many-body states can be thought of, in a mean field sense, as arising from the spectral gaps between the *single-particle* Landau levels of quasi-particles known as "composite fermions." Thus in the composite fermion picture, the fractional quantum Hall effect could be viewed as a single-particle phenomenon — the integer quantum Hall effect of composite fermions.[2] The composite fermion point of view has proved to be fruitful theoretically,[2-4] and much experimental evidence has accumulated supporting its validity.[5-18]

Recently a model of single-particle composite fermion edge states has been proposed[19] that generalizes the very successful edge state theories of transport in the integer quantum Hall effect[20-23] to the fractional regime. This model explains[19] the fractionally quantized Hall conductances that are observed experimentally. It also explains[19] the results of transport experiments on Hall bars with potential barriers, including experiments[24,25] that involve selective population and detection of fractional edge channels. In this model, the composite fermion edge states have unconventional properties[19] but the quasi-particles that populate them are assumed to form a Fermi liquid, in contrast to what is assumed in chiral Luttinger liquid edge state models.[26]

Whether the edge states of *real* two-dimensional electron systems in the fractional quantum Hall regime are Fermi or Luttinger liquids is an important unresolved question. It can only be answered definitively by comparing the predictions of Fermi and Luttinger liquid models with experiments.

Experiments on selective population and detection of fractional edge channels indicate that more than one conducting mode is present at the edge in $\nu = 1$ systems,[24,25] where $\nu$ is the bulk Landau level filling fraction. This is in agreement with the composite fermion edge state model,[19] but contrary to the assumption of the published chiral Luttinger liquid work (that is based on abrupt edge potentials)[26] that there should be only one mode present at a $\nu = 1$ edge. The observation[27-30] of intermediate fractional plateaus (between $G = \nu e^2/h$ and $G = 0$) in the two-terminal conductance $G$ of constrictions (or point contacts) has also not been explained by the Luttinger liquid models, but is explained by the composite fermion edge state model, as will be shown below. However, it is not clear whether the Luttinger liquid models that are available at present do not explain these experiments for fundamental reasons or simply because they do not include significant physical features of the real experimental devices.

Experiments measuring the temperature dependence of transport through semiconductor nanostructures connecting 2D electron gases in the fractional regime should in principle be able to discriminate between Fermi and Luttinger liquid models more convincingly. However, so far the results of such experiments have been mixed: Measurements of quantum transmission between fractional edge states via a *disordered* constriction[31] have yielded results in agreement with the Luttinger liquid models. But whether these measurements are really evidence of a Luttinger liquid is unclear because the mechanism by which electrons are transmitted through a disordered constriction in the fractional quantum Hall regime is not understood, which may leave room for other possible interpretations of the data.[32] In order to try to address this issue, experiments have recently been performed on a related but cleaner system — the temperature dependence of Aharonov-Bohm conductance resonances of a ballistic constriction containing an artificial impurity (or antidot) has been measured in the fractional regime. These experiments yielded results in agreement with a Fermi liquid picture of fractional edge states.[29,33] However, no Fermi liquid theory of transport through these devices in the fractional quantum Hall regime has been available for a detailed comparison. Such a theory based on the composite fermion edge state model is presented in this article. Although in this paper I focus on transport through simple adiabatic and non-adiabatic constrictions and constrictions containing artificial impurities, the theoretical framework developed here should be useful in studying many different semiconductor nanostructures.

In Section 2, I summarize briefly the relevant features of the model of composite fermion edge states. In Section 3 a theory of composite fermion edge state transport through adiabatic and non-adiabatic constrictions in the fractional regime is developed. Adiabatic constrictions are shown to exhibit a series of fractionally quantized conductance plateaus as their geometry is varied, in agreement with experiments. Interestingly, the temperature dependence of the conductance of constrictions predicted by the composite fermion theory is found to differ from what one might naively expect on the basis of ordinary Fermi liquid phenomenology. The difference, although small, should in principle be detectable by careful measurements. In Section 4, I generalize the results obtained in Section 3 to constrictions containing artificial impurities, focussing particularly on the regime in which the recent experiments[29,33] were performed. It is predicted that these devices should under suitable conditions exhibit behavior in the fractional regime that is qualitatively similar to that which they show in the integer regime, including transmission and reflection resonances and Aharonov-Bohm oscillations with beats and other effects. The temperature dependence predicted for these features is close but not identical to that expected for an ordinary Fermi liquid, and is consistent with the recent experiments.[29,33] The conclusions are presented in Section 5.

## 2. The Composite Fermion Model of Fractional Edge States

In the Chern-Simons version of composite fermion theory, a singular gauge transformation is performed which attaches a tube of fictitious gauge flux carrying an even number of flux quanta to each electron. [2-4] The electrons together with the attached flux tubes obey Fermi statistics and are referred to as "composite fermions."[2] In mean field theory the interactions between composite fermions that are due to the vector potentials associated with the tubes of fictitious gauge flux, are replaced by interactions with a fictitious average magnetic field and a fictitious electric field. The fictitious magnetic field [2-4] is $\mathbf{B}_g = -mn\hat{\mathbf{B}}h/e = -m\nu\mathbf{B}$ where $m$ is the number of quanta of gauge flux carried by each composite fermion, $n$ is the two-dimensional electron density, $\nu = nh/(eB)$ is the Landau level filling parameter, and $\hat{\mathbf{B}}$ is the unit vector in the direction of the true magnetic field $\mathbf{B}$. Thus the composite fermions interact with a total effective magnetic field $\mathbf{B}_{eff} = \mathbf{B} + \mathbf{B}_g$. The fictitious electric field is[34] $\mathbf{E}_g = -(\mathbf{J} \times \hat{\mathbf{B}})mh/e^2$, where $\mathbf{J}$ is the two-dimensional electric current density.

In the composite fermion model of edge states[19] it is assumed that the electron density is slowly varying with position and that, for local Landau level filling fractions in the vicinity of $1/m$, the composite fermion Landau level energies behave qualitatively like

$$\varepsilon_{m,r} = \left(r + \frac{1}{2}\right)\hbar e|B_{eff}|/m^* + W. \quad (1)$$

Here $r = 0, 1, 2,...$ and $W$ is the position-dependent composite fermion effective potential energy. $m^*$ is the composite fermion effective mass. Equation (1) is a good description of the composite fermion Landau level structure in uniform systems[2] and yields edge states that propagate in the direction consistent with experiments.[35] The assumed behavior of the composite fermion Landau levels near an edge is depicted schematically in Fig.1. The effective magnetic field $\mathbf{B}_{eff}$ well away from the edge is parallel to the real magnetic field in Fig.1(a) and anti-parallel in Fig.1(b). The apex of each "fan" of energy levels occurs where and $\mathbf{B}_{eff} = 0$ for some even integer $m$; there according to equation (1) $\varepsilon_{m,r} = W$. The different types of composite fermion edge states are labelled I, II and III. Type I edge states are silent modes in the sense that the currents that they carry are independent of the electrochemical potential at the edge, provided that quasi-equilibrium conditions prevail there.[19] However, Landau levels of types II and III carry net currents of magnitude $|e\Delta\mu^*/h|$ where $\Delta\mu^*$ is the difference between the composite fermion effective electrochemical potentials at opposite edges of the sample. From the gauge invariance of the electron charge density, it follows[19] that the effective electrochemical potentials $\mu^*$ of composite fermions at the edges of a current-carrying Hall bar differ, in general, from the corresponding electron electrochemical potentials $\mu$, and that

$$\mu_\alpha^* - \mu_\beta^* = \mu_\alpha - \mu_\beta + \int_\beta^\alpha e\mathbf{E}_g \cdot d\mathbf{r} = \mu_\alpha - \mu_\beta - \int_\beta^\alpha (mh/e)\mathbf{J}\cdot\hat{\mathbf{B}} \times d\mathbf{r}, \quad (2)$$

where $\alpha$ and $\beta$ label *different* edges of the Hall bar. Equation (2) is important because while the composite fermion currents are controlled by the *composite fermion* electrochemical potentials $\mu^*$, the voltages measured experimentally at the contacts reflect directly the *electron* electrochemical potentials $\mu$. Thus equation (2) connects the composite fermion theory with experiments.

## 3. Composite Fermion Edge Transport Through Constrictions

Perhaps the simplest semiconductor nanostructure is a constriction in a two-dimensional electron gas (2DEG).[37] Such systems can support many different configurations of composite fermion edge states in the fractional quantum Hall regime. In this Section we will analyze three important classes of configurations that are shown schematically in Fig.2.

In the upper panels of Fig.2 the black areas are depleted of electrons. The different directed curves show different *groups* of type I and III composite fermion Landau levels (that derive from the bulk value of *m*) where they pass through the Fermi energy $E_F$. The thinner curves indicate where $B_{eff} = 0$ for the bulk value of *m*. In Fig.2(b) and Fig.2(c) there are regions occupied by type II modes below the Fermi level; these regions are so labelled. The inter-edge state scattering processes that are important are represented by the dashed lines. The scattering is represented here as partial reflection of edge states at the constrictions; equivalent results can of course be obtained by representing the scattering as partial transmission instead. Each of the lower panels shows schematically an example of the composite fermion Landau level energy spectrum (for the bulk value of *m*) as a function of position along a line running from left to right through the center of the constriction shown in the panel above.

**Case (a)**

Consider now case (a) of Fig.2, where the effective magnetic field $\mathbf{B}_{eff}$ is parallel to the real magnetic field $\mathbf{B}$ both in the bulk and at the center of the constriction.

*The adiabatic limit*, where there is no inter-edge state scattering, can be treated by methods similar to those used to study transport across adiabatic potential barriers in Ref. 19. At zero temperature, the net current flowing through the constriction can be represented as the sum of the contributions of the type III occupied states flowing through the constriction where the constriction is narrowest:

$$I_{\text{adiabatic}} = -\frac{e}{h}\sum_r \int \frac{\partial \varepsilon_{m,r}}{\partial k} dk = -\frac{e}{h}\sum_r \int d\varepsilon_{m,r} \qquad (3)$$

where *k* is the wave vector in the direction through the constriction and *r* labels the occupied type III Landau levels at the center of the constriction. If $\mu_1^*$ and $\mu_2^*$ are the composite fermion edge electrochemical potentials as indicated in Fig.2(a), this yields

$$I_{\text{adiabatic}} = -p\frac{e}{h}\left(\mu_1^* - \mu_2^*\right), \qquad (4)$$

where *p* is the number of composite fermion Landau levels populated at the center of the constriction. That is, each composite fermion Landau level that is transmitted through the constriction contributes a current $-e(\mu_1^* - \mu_2^*)/h$. This result resembles Büttiker-Landauer theory.[23] There is the important difference, however, that the electrochemical potentials that appear here are those of composite fermions, not electrons. Since the experimentally measured voltages of the contacts are given by the *electron* electrochemical potentials,[19] the composite fermion electrochemical potentials must be eliminated in terms of the electron electrochemical potentials in order to obtain results that can be compared with experiments. This can be done by using equation (2), which takes the form

$$\mu_1^* - \mu_2^* = \mu_1 - \mu_2 + m\frac{h}{e}I, \tag{5}$$

where $I$ is the net current flowing through the constriction. This yields

$$G_{\text{adiabatic}} = -e\frac{I_{\text{adiabatic}}}{\mu_1 - \mu_2} = \frac{e^2}{h}\frac{p}{mp+1} \tag{6}$$

for the two-terminal conductance of the constriction. Notice that the conductance (6) is fractional for $m \geq 2$, despite the fact that the composite fermion charge is integer ($-e$), because the composite fermion electrochemical potentials differ from those of electrons. Equation (6) also means that as an adiabatic constriction is pinched down (and $p$ changes), its two-terminal conductance should exhibit a series of fractional plateaus. This is in agreement with experiments[27-30] but differs qualitatively from the results that have been obtained to date[26] using Luttinger liquid models[26] of transport through constrictions. (These Luttinger liquid models[26] do not predict the occurrence of any intermediate conductance plateaus between $G = \nu_{\text{bulk}}e^2/h$ and $G = 0$.)

The above calculation can be generalized to the non-adiabatic case as follows. Since the energetics of the type III composite fermion Landau levels in Fig.2(a) is qualitatively similar to the Landau level structure of electrons in the integer quantum Hall regime, it is legitimate to apply Büttiker-Landauer theory to this case *within the composite fermion picture*. For the current through the constriction this yields

$$I = -\frac{e}{h}\sum_{i,j}T_{ij}^*\left(\mu_1^* - \mu_2^*\right) \tag{7}$$

instead of the adiabatic result (4). Here $T_{ij}^*$ is the probability that a composite fermion incident on the constriction (from the source contact) at the Fermi level in a type III state $j$ is transmitted through the constriction to emerge in state $i$. The relevant scattering processes are represented by the dashed line in Fig.2(a). To eliminate ($\mu_1^*$-$\mu_2^*$) once again in terms of the corresponding electron electrochemical potentials, the path of integration in equation (2) should begin and end well outside of the scattering region (for example at points $\beta$ and $\alpha$ in Fig.2(a)) where the edge electrochemical potentials are well defined. Then equation (5) still holds, and combining it with equation (7) one finds

$$G_0 = -e\frac{I}{\mu_1 - \mu_2} = \frac{e^2}{h}\frac{\sum_{i,j}T_{ij}^*}{m\sum_{i,j}T_{ij}^* + 1} \tag{8}$$

for the two-terminal conductance of the constriction.[38] Notice that if $p$ composite fermion Landau levels are transmitted perfectly through the constriction and all others are reflected then $\Sigma_{ij} T_{ij}^* = p$ and equation (8) reduces to the adiabatic result (6) as it should.

The above treatment applies at zero temperature. The effects of the $kT$ broadening of the Fermi surface on the conductance at finite temperatures can be evaluated as follows. In the linear response regime, the *thermally averaged* current $<I>$ at finite temperatures can be obtained by replacing $\Sigma_{ij}T_{ij}^*$, the composite fermion transmission probability at the Fermi energy in equation (7), by the energy integral $-\Sigma_{ij}\int d\varepsilon\, \partial f^*/\partial\varepsilon T_{ij}^*(\varepsilon)$. Here $f^*=(e^{(\varepsilon-\mu^*)/kT}+1)^{-1}$ is the composite fermion Fermi function and $T_{ij}^*(\varepsilon)$ is the energy-dependent composite Fermion transmission probability.

For $|\mu_1^* - \mu_2^*| \ll kT$ this yields

$$\langle I \rangle = \frac{e}{h}\left(\mu_1^* - \mu_2^*\right)\sum_{i,j}\int T_{ij}^*(\varepsilon)\frac{\partial}{\partial \varepsilon}f^* d\varepsilon \tag{9}$$

Taking the thermal average of equation (2) yields

$$\mu_1^* - \mu_2^* = \mu_1 - \mu_2 + m\frac{h}{e}\langle I \rangle. \tag{10}$$

Finally, combining equations (9) and (10) yields the result

$$G_T = -e\frac{\langle I \rangle}{\mu_1 - \mu_2} = \frac{e^2}{h}\frac{-\sum_{i,j}\int T_{ij}^*(\varepsilon)\frac{\partial}{\partial \varepsilon}f^* d\varepsilon}{1 - m\sum_{i,j}\int T_{ij}^*(\varepsilon)\frac{\partial}{\partial \varepsilon}f^* d\varepsilon} \tag{11}$$

for the finite temperature two-terminal conductance of the constriction.

This is an interesting result since it shows that the standard relation

$$G_T^F = -\int G_0(\varepsilon)\frac{\partial f}{\partial \varepsilon}d\varepsilon \tag{12}$$

between the zero temperature conductance $G_0(\varepsilon)$ (at Fermi energy $\varepsilon$) and the finite temperature conductance $G_T^F$ of non-interacting fermion systems does *not* apply to composite fermion systems.

To make the implications of this difference clear, let us consider the temperature dependence of the conductance of an *adiabatic* constriction in the above regime. In that case the total transmission probability at composite fermion energy $\varepsilon$ is given by

$$\sum_{i,j}T_{ij}^*(\varepsilon) = p(\varepsilon) \tag{13}$$

where $p(\varepsilon)$ is an integer equal to the number of composite fermion Landau levels that are below the energy $\varepsilon$ at the center of the constriction. Assuming, for simplicity that $kT$ is much smaller than the composite Fermion transverse energy level spacing in the constriction, equation (11) then yields

$$G_T \approx \frac{e^2}{h}\frac{(r-1)e^{-\beta\mu'} + r}{(mr - m + 1)e^{-\beta\mu'} + mr + 1} \tag{14}$$

where $\beta = 1/kT$; $\mu'$ is the composite fermion electrochemical potential *measured from the bottom of the $r^{th}$ composite fermion Landau level at the center of the constriction* and is assumed to be small. On the other hand, under the same conditions, equation (12) yields

$$G_T^F \approx \frac{e^2}{h}\left(\frac{(r-1)e^{-\beta\mu'}}{mr - m + 1} + \frac{r}{mr + 1}\right)(e^{-\beta\mu'} + 1)^{-1}. \tag{15}$$

Expressions (14) and (15) are plotted in Fig.3(a) in the vicinity of the cross-over between the $e^2/3h$ and $2e^2/5h$ two-terminal conductance plateaus of the constriction, which occurs for $r = m = 2$. The solid line is $G_T$, the prediction of the composite Fermion theory. The dashed line is $G_T^F$, the result of naively applying equation (12), the usual result for ordinary Fermi liquids. The conductance $G_T$ predicted by the composite fermion theory is somewhat larger than $G_T^F$ throughout the transition.

The difference between the two curves can also be described as a rigid shift by $\ln((mr+1)/(mr-m+1))$ in $\mu'/kT$.

**Case (b)**

In the case shown in Fig.2(b), $\mathbf{B}_{\text{eff}}$ is anti-parallel to $\mathbf{B}$ both in the bulk and at the center of the constriction. Since the type II Landau levels play a key role in the conductance in this case although they are entirely below the Fermi level and separated from it by a gap (see the lower panel of Fig.2(b)), this situation is qualitatively different from those that occur in Büttiker-Landauer theory.

In the adiabatic regime, where there is no inter-edge state scattering, applying equation (3) to the type II Landau levels at the narrowest part of the constriction, yields a current flowing through the constriction that is given by

$$I_{\text{adiabatic}} = p\frac{e}{h}\left(\mu_1^* - \mu_2^*\right), \tag{16}$$

where $p$ is the number of composite fermion Landau levels populated at the center of the constriction. (In the adiabatic case, the type I edge states that also pass through the constriction do *not* contribute to the net current response of the constriction. This follows from the same argument that was used in Ref. 19 to demonstrate that type I edge states are silent in the wide Hall bars.) Equation (5) applies to the present case as well, and using it to eliminate the composite fermion electrochemical potentials from (16) in terms of the electron electrochemical potentials yields

$$G_{\text{adiabatic}} = -e\frac{I_{\text{adiabatic}}}{\mu_1 - \mu_2} = \frac{e^2}{h}\frac{p}{mp-1} \tag{17}$$

for the adiabatic two-terminal conductance of the constriction. Thus we find once again that the two-terminal conductance of the constriction should exhibit a series of fractional plateaus as the constriction is narrowed and $p$ changes, consistent with experiments.[27-30]

Consider now the case of Fig.2(b) in the *non-adiabatic* regime at zero temperature. At zero temperature, the type II composite fermion Landau levels shown in the lower panel of Fig.2(b) are completely filled states below the Fermi level. Thus scattering processes involving them do not affect the current flowing through the constriction. The current that they carry is thus still given by equation (16). However, the type I Landau levels in the constriction cross the Fermi level and therefore the scattering processes that involve them can be important. In the adiabatic regime, the type I modes are "silent" in the sense that they do not contribute to the change in the current through the constriction when the electrochemical potentials $\mu_1$ and $\mu_2$ change. Thus the *total* current that they carry through the constriction in the absence of scattering must be zero, and consequently their net effect on the conductance of the constriction is *entirely* due to the scattering processes. The current through the constriction can then be obtained as follows: Let $R_{ij}^*$ be the probability that a composite fermion at the Fermi level in the type I edge state $j$ incident on the constriction from the source contact is reflected at the constriction by being scattered into edge state $j$. Then $\Sigma_{ij}R_{ij}^*$ is the total reflection probability for composite fermions in the type I channels at the constriction. Let $q$ be the number of type II composite fermion Landau levels occupied in the bulk, well away from the constriction. The total current through the constriction is then

$$I = -\frac{e}{h}\left(-q - \sum_{i,j}R_{ij}^*\right)\left(\mu_1^* - \mu_2^*\right); \tag{18}$$

this is the generalization of equation (16) to the non-adiabatic regime. Equation (5) still applies, and combining it with (18) yields

$$G_0 = -e\frac{I}{\mu_1 - \mu_2} = \frac{e^2}{h}\frac{q + \sum_{i,j} R_{ij}^*}{m\left(q + \sum_{i,j} R_{ij}^*\right) - 1} \tag{19}$$

for the zero temperature two-terminal conductance of the constriction.

*In the adiabatic limit* each of the type I edge states is either perfectly transmitted through the constriction or perfectly reflected. $\Sigma_{ij}R_{ij}^*$ is then simply the number of the type I edge states that is perfectly reflected, and the number $p$ of type II Landau levels at the center of the constriction equals the number $q$ present in the bulk plus the number of type I modes that are reflected. Therefore $p = q + \Sigma_{ij}R_{ij}^*$ and (19) reduces to the adiabatic result (17), as it should.

The temperature-dependent two-terminal conductance of the constriction for the case in Fig.2(b) in the non-adiabatic regime may be calculated as follows. Referring to Fig.4, the net current through the constriction is given by

$$\langle I \rangle = -\frac{e}{h}\sum_{i,j}\int_{\varepsilon_\beta}^{\varepsilon_\alpha} T_{ij}^*(\varepsilon) f^*(\varepsilon - \mu_1^*) d\varepsilon - \frac{e}{h}\sum_{i,j}\int_{\varepsilon_\alpha}^{\infty} T_{ij}^*(\varepsilon)\Big(f^*(\varepsilon - \mu_1^*) - f^*(\varepsilon - \mu_2^*)\Big) d\varepsilon \tag{20}$$

where $T_{ij}^*(\varepsilon)$ is the probability that a particle in channel $j$ incident on the constriction from source contact in Fig.2(b) at energy $\varepsilon$ is transmitted through the constriction into channel $i$ on the other side. We now assume (generalizing the zero temperature *ansatz* introduced in Ref. 19 to finite temperatures) that, at fixed $B$ and temperature, the difference between the composite fermion edge electrochemical potential and the bottom-of-the-band energy where $B_{\text{eff}}=0$, depends only on the local electron density where $B_{\text{eff}}=0$. Since this density is set by the condition $\nu=1/m$ where $m$ is an even integer, it follows that in Fig.2(b) $\mu_1^*-\varepsilon_\beta =\mu_2^*-\varepsilon_\alpha$. Assuming also that $kT << \mu_1^*-\varepsilon_\alpha$ and that $|\mu_1^*-\mu_2^*| <<kT$ equation (20) becomes

$$\langle I \rangle = \frac{e}{h}\left(N + \sum_{i,j}\int_{\varepsilon_\alpha}^{\infty} d\varepsilon T_{ij}^*(\varepsilon)\frac{\partial}{\partial \varepsilon}f^*\right)\left(\mu_1^* - \mu_2^*\right) \tag{21}$$

Where $N$ can be thought of as the total number of channels present in Fig.2(b) in the energy range $\varepsilon_\beta < \varepsilon < \varepsilon_\alpha$; more formally, $N=\Sigma_{ij}T_{ij}^*(\varepsilon)$ which is independent of $\varepsilon$ in this range, as follows from the unitarity of the scattering matrix since all of the channels at the same edge propagate in the same direction in the present model.[19] Equation (10) applies here as well and combining it with (21) yields

$$G_T = -e\frac{\langle I \rangle}{\mu_1 - \mu_2} = \frac{e^2}{h}\frac{N + \sum_{i,j}\int_{\varepsilon_\alpha}^{\infty} d\varepsilon T_{ij}^*(\varepsilon)\frac{\partial}{\partial \varepsilon}f^*}{m\left(N + \sum_{i,j}\int_{\varepsilon_\alpha}^{\infty} d\varepsilon T_{ij}^*(\varepsilon)\frac{\partial}{\partial \varepsilon}f^*\right) - 1} \tag{22}$$

for the finite temperature two-terminal conductance of the constriction.

In the zero temperature limit equation (22) reduces to equation (19) as expected since

$$\lim_{T \to 0}\left(N + \sum_{i,j}\int_{\varepsilon_\alpha}^{\infty} d\varepsilon T_{ij}^*(\varepsilon)\frac{\partial}{\partial \varepsilon}f^*\right) = N - \sum_{i,j} T_{ij}^* = q + \sum_{i,j} R_{ij}^* \tag{23}$$

where $R_{ij}^*$ on the right hand side and $T_{ij}^*$ in the center are evaluated at the Fermi level.

If the constriction is perfectly adiabatic, each mode is either perfectly transmitted or perfectly reflected by it at the Fermi level. Hence $\Sigma_i T_{ij}^* = 1$ or 0 for each mode. If we also assume that $kT$ is much smaller than the energy spacing of the type II Landau levels at the center of the constriction, equation (22) reduces in the adiabatic limit to

$$G_T \approx \frac{e^2}{h} \frac{r + e^{-\beta\mu'}(r-1)}{(mr-1) + e^{-\beta\mu'}(mr-m-1)} \tag{24}$$

where $\mu'$ is the composite fermion electrochemical potential *measured from the top of the $r^{th}$ composite fermion Landau level at the center of the constriction* and is assumed to be small. Under the same conditions, equation (12), the corresponding standard Fermi liquid result, becomes

$$G_T^F \approx \frac{e^2}{h}\left(\frac{(r-1)e^{-\beta\mu'}}{mr-m-1} + \frac{r}{mr-1}\right)(e^{-\beta\mu'} + 1)^{-1}. \tag{25}$$

The transition between the $2e^2/3h$ and $3e^2/5h$ conductance plateaus (the case $m=2$, $r=3$) as predicted by expressions (24) and (25) is shown in Fig.3(b). The solid line is $G_T$, the prediction of the composite fermion theory. The dashed line is $G_T^F$. It may at first seem surprising that both $G_T$ and $G_T^F$ decrease with increasing $\mu'$ in Fig.3(b), in contrast to the behavior seen Fig.3(a). However, the reason for this behavior is that $\mu'$ is defined *relative* to the energies of the composite fermion Landau levels. As $\mu'$ increases more composite fermion Landau levels become populated at the center of the constriction (see Fig.4). For $\mathbf{B}_{eff}$ anti-parallel to $\mathbf{B}$, this implies a *decrease* of the two-terminal conductance of the constriction, just as increasing the number of occupied composite fermion Landau in a wide Hall bar results in a decrease of its Hall conductance if $\mathbf{B}_{eff}$ is anti-parallel to $\mathbf{B}$. The difference between $G_T$ and $G_T^F$ is once again a rigid shift in $\mu'/kT$, in this case by $\ln((mr-1)/(mr-m-1))$.

**Case (c)**

The case in Fig.2(c) where the effective magnetic field $\mathbf{B}_{eff}$ is anti-parallel to the real magnetic field $\mathbf{B}$ in the bulk but parallel to $\mathbf{B}$ at the center of the constriction is in general more complicated than those discussed above. However, it simplifies greatly if one supposes that the only modes that are not totally reflected by the constriction are the type I modes that become type III modes near the center of the constriction, as indicated in the figure. This is a not unreasonable assumption for a constriction that is smooth enough to be nearly adiabatic. Under these conditions the analysis for case (a) above can be applied directly to the type III modes in Fig.2(c), and the results obtained for the two-terminal conductance of the constriction in case (a) hold here as well provided that $T_{ij}^*$ is interpreted as the transmission matrix for the type III modes in Fig.2(c).

An interesting implication of the above results (that is evident in Fig.3) is that the finite temperature conductance $G_T$ of an adiabatic constriction in the transition between plateaus is predicted by the present composite fermion theory to be larger than the naive Fermi liquid prediction $G_T^F$ if $\mathbf{B}_{eff}$ is parallel to $\mathbf{B}$ at the center of the constriction, but smaller than $G_T^F$ if $\mathbf{B}_{eff}$ is anti-parallel to $\mathbf{B}$ there. Experiments designed to detect this effect would be of interest.

## 4. Transport Through a Constriction Containing an Antidot

The above theory can be generalized to treat conduction through more complex nanostructures in the fractional quantum Hall regime. An important example is a constriction containing an artificial impurity or "anti-dot." Such devices have been the subject of experiments in both the integer[40,41] and fractional[27,28,29,33] quantum Hall regimes. They are of considerable fundamental interest because experiments performed on them should help decide whether edge states in the fractional quantum Hall regime are Fermi or Luttinger liquids. A case that has been studied experimentally in the fractional regime is that in which the effective magnetic field is antiparallel to the real magnetic field in the bulk and parallel to the real magnetic field in the two conducting channels between the anti-dot and the walls of the constriction, and each of the channels transmits the same number of modes.[27,28,29,33] This case is represented schematically in Fig.5(a), and will be discussed here. (The results obtained can also be applied directly to the case where the effective magnetic field is parallel to the real magnetic field in the bulk as well as in the conducting channels).

The notation in Fig.5(a) is the same as in Fig.2. The antidot is the black region at the center of the figure. It is depleted of electrons. The dashed lines indicate the most important composite fermion edge state scattering processes at the Fermi energy. Since the effective magnetic field is parallel to the real magnetic field in the region between the thin curves (which indicate where $\mathbf{B}_{\text{eff}} = 0$ for the bulk value of $m$), this situation is somewhat analogous to case (c) of the preceding Section. In analyzing it I will assume for simplicity that the only edge state scattering processes that are significant are those denoted **c**, **d**, **e**, **f**, **g** and **h** in the figure. Then the solution of the problem becomes formally similar to that for the non-adiabatic constrictions in Fig.2(a) and Fig.2(c), and equations (5) and (7) - (11) are also valid for the system in Fig.5(a). However in this case $T_{ij}^*$ should be interpreted as the probability that a composite fermion in edge state $j$ at the Fermi energy that is incident from the source onto the constriction containing the antidot, is transmitted through the region containing the constriction *and antidot* to emerge in edge state $i$ on the other side.

Topologically similar edge state and scattering patterns that occur in constrictions and quantum wires containing an antidot in the *integer* regime have been analyzed theoretically in Ref. 42 and analytic expressions for the multi-channel transmission probability $\Sigma_{ij}T_{ij}$ have been obtained there for a number of cases. These analytic results can be used directly in the present composite fermion calculations because the *analytic forms* of $\Sigma_{ij}T_{ij}$ derived in Ref. 42 depend only on the pattern of edge states and inter-edge state scattering and not on whether the particles involved are ordinary electrons in the integer regime or composite fermions in the fractional regime. The functional form of $\Sigma_{ij}T_{ij}$ in Ref. 42 is the same as that of $\Sigma_{ij}T_{ij}^*$ for the corresponding scattering pattern in the present case. In the integer regime the zero temperature two-terminal conductance of the device is given by $G_{\text{int}} = e^2/h\, \Sigma_{ij}T_{ij}$ while the corresponding result (8) of the composite fermion theory in the fractional regime is $G_0 = e^2/h\, \{\Sigma_{ij}T_{ij}^*/(m\Sigma_{ij}T_{ij}^*+1)\}$. Since $G$ is a monotonically and smoothly increasing function of $\Sigma_{ij}T_{ij}^*$ and the functional dependence of $\Sigma_{ij}T_{ij}^*$ on the model parameters is the same as that of $\Sigma_{ij}T_{ij}$, it follows that all of the qualitative transport phenomena predicted in Ref. 42 for the integer regime should also occur in the fractional regime. That, is the reflection and transmission resonances, the beats due to quantum interference between different edge states and the abrupt phase changes of the conductance oscillations discussed in Ref. 42 should also all occur under suitable conditions in the fractional regime depicted in Fig.5(a). One can think of this correspondence between the results for the integer and fractional regimes as a rescaling of the dimensionless conductance $g = G_0/(e^2/h)$ at zero temperature according to $g \rightarrow g/(mg+1)$. Although this scaling function is not linear it is smooth and thus the line shapes of var-ious conductance features in the fractional regime closely resemble the corresponding ones of the integer regime.

The above analysis indicates that Jain's proposition[2] that the fractional quantum Hall effect is the integer quantum Hall effect for composite fermions should also apply to the *breakdown* of the quantum Hall effect in non-adiabatic nanoscale devices with scattering between edge states.

As a simple example, let us suppose that the only edge state scattering that is significant in Fig.5(a) is that at **e** and at **h** (the regime of resonant reflection in Ref. 42), and that two composite fermion Landau levels derived from the bulk value of *m* are populated at the center of each channel between the antidot and the walls of the constriction. Suppose also, for simplicity, that composite fermions belonging to one of these Landau levels are transmitted perfectly through each channel while those of the other Landau level are transmitted with probabilities $T_e$ and $T_h$ through the channels at **e** and at **h** in Fig.5(a) respectively. The assumed composite fermion Landau level structure and inter-edge state scattering in the channel at **e** are depicted schematically in Fig.5(b). Applying the results of Ref. 42 to this case yields

$$\sum_{i,j} T_{ij}^* = 2 - \frac{(1-T_e)(1-T_h)}{T_e T_h - 2\sqrt{T_e T_h}\cos(\Phi) + 1}. \tag{26}$$

where $\Phi$ is the Aharonov-Bohm phase (including the scattering phase shifts) associated with the outer composite fermion orbit of the antidot. Note that the composite fermion charge is $-e$. This implies that the Aharonov-Bohm period for the orbit is close to that for non-interacting electrons since the effective magnetic field is the same as the true magnetic field inside the antidot which is depleted of electrons. Some typical results for $G_0$ (obtained from equations (8) and (26)) for $m = 2$, and the corresponding results for $G_{int}$ in the integer regime (for the same values of the model parameters $T_e$ and $T_h$) are shown in Fig.6(a) and Fig.6(b) respectively. Notice that although the corresponding conductance line shapes in the integer and fractional regimes are not identical, they are very similar, consistent with the above scaling argument. The features in Fig.6 are reflection resonances in the conductance. If scattering processes at **c**, **d**, **f** and **g** in Fig.5(a) are included in the calculation (using the analytic expressions for $\Sigma_{ij}T_{ij}$ given in Ref. 42) then the other phenomena discussed in Ref. 42, such as transmission resonances and beats, also appear in the conductance.

The temperature dependence of the conductance resonances that is due to the *kT* broadening of the composite fermion Fermi surface can be calculated for the system shown in Fig.5(a) by using equation (11). Representative results obtained for the simple model of resonant reflection described by equation (26) are shown in Fig.7(a), where it is assumed that $T_e$ and $T_h$ are independent of energy and the Aharonov-Bohm phase $\Phi$ depends linearly on energy ($\Phi \sim c\varepsilon$) in the energy range of interest near the Fermi level. $G_0$ is the zero temperature conductance, $G_T$ is the conductance calculated using the present composite fermion theory (equation (11)) at a temperature $T=0.05\pi/(kc)$. $G_T^F$ is the prediction (12) of ordinary Fermi liquid phenomenology at the same temperature. For comparison, the corresponding results for a simple transmission resonance with only the scattering processes **c** and **f** in Fig.5(a) active and only one composite fermion Landau level transmitted through each of channels e and h, is shown in Fig.7(b). In this case equation (26) is replaced[42] by

$$\sum_{i,j} T_{ij}^* = 1 + \frac{(1-T_c)(1-T_f)}{T_c T_f - 2\sqrt{T_c T_f}\cos(\Phi) + 1}, \tag{27}$$

where $1-T_c$ and $1-T_f$ are the probabilities of inter-edge state scattering at **c** and **f**.

Notice that although the prediction $G_T$ of composite fermion theory differs from that of Fermi liquid phenomenology $G_T^{\mathrm{F}}$, the difference is small and very careful measurements would be needed to detect it experimentally. Another interesting feature of Fig.7 is that the finite temperature reflection resonance predicted by composite fermion theory is weaker than that predicted by naive Fermi liquid theory whereas the opposite is true of the transmission resonance.

Since the temperature dependence of Aharonov-Bohm resonances predicted by the present composite fermion theory is very close to that predicted by the naive Fermi liquid phenomenology, which is known to provide a good empirical description of the presently available experimental results,[29,33] it is clear that the composite fermion theory is able to explain the experimental Aharonov-Bohm data that has been obtained to date very well.

## 5. Conclusions

In this article a composite fermion (Fermi liquid) edge state theory of transport through semiconductor nanostructures in the fractional quantum Hall regime has been presented. The theory is in good agreement with recent experiments on clean constrictions[27-30] and constrictions containing artificial impurities.[29,33] The temperature dependence of the conductance of these devices is predicted to be close to (although not exactly the same as) that predicted by ordinary Fermi liquid phenomenology, and is in agreement with experiments. The differences between the predictions of the present composite fermion theory and ordinary Fermi liquid behavior should be experimentally observable but their observation will require very careful measurements on high quality devices.

Acknowledgments — I would like to thank B. L. Johnson, C. J. B. Ford, A. Sachrajda, J. Frost, C. -T. Liang, C. Barnes, A.M. Chang, P. Coleridge and M. Geller for interesting discussions. This work was supported by NSERC, Canada.

**Figure Captions:**

**Fig.1** Schematic of the composite fermion Landau level structure near an edge. The effective magnetic field $\mathbf{B}_{\text{eff}}$ well away from the edge is parallel to the real magnetic field in (a) and anti-parallel in (b). The apex of each "fan" of energy levels occurs where and $\mathbf{B}_{\text{eff}} = 0$ for some even integer $m$; there $\varepsilon_{m,r} = W$. The different types of composite fermion edge states are labelled I, II and III.

**Fig.2** Some possible composite fermion Landau level configurations in constrictions. In the upper panels the black areas are depleted of electrons; the different directed curves indicate where different groups of type I and III composite fermion Landau levels pass through the Fermi energy $E_{\text{F}}$, and the thinner curves indicate where $B_{\text{eff}} = 0$ for the principal value of $m$. The dashed lines indicate scattering between composite fermion Landau levels. Regions containing type II composite fermion Landau levels are indicated. Each of the lower panels show schematically an example of the composite fermion Landau level energies (for the principal value of $m$) as a function of position along a line running from left to right through the center of the constriction in the panel above. (a) $\mathbf{B}_{\text{eff}}$ is parallel to the real magnetic field $\mathbf{B}$ both in the bulk and at the center of the constriction. (b) $\mathbf{B}_{\text{eff}}$ is anti-parallel to $\mathbf{B}$ both in the bulk and at the center of the constriction. (c) $\mathbf{B}_{\text{eff}}$ is anti-parallel to $\mathbf{B}$ in the bulk and parallel to $\mathbf{B}$ at the center of the constriction.

**Fig.3** Transitions between fractional plateaus of the two-terminal conductance of an adiabatic constriction at finite temperatures. (a) $\mathbf{B}_{\text{eff}}$ is parallel to $\mathbf{B}$ at the center of the constriction. (b) $\mathbf{B}_{\text{eff}}$ is anti-parallel to $\mathbf{B}$ at the center of the constriction. The solid lines are $G_T$ the prediction of the present composite fermion theory; the dashed lines are $G_T^{\text{F}}$ the result of naive Fermi liquid theory.

**Fig.4** Schematic drawing of the composite fermion Landau level structure showing the type I and II modes in the constriction in Fig.2(b) together with the composite fermion electrochemical potentials (outside of the constriction) of the edges that feed the constriction.

**Fig.5** (a) Composite fermion edge state configuration for a constriction with an antidot (the black central region) in the regime where the effective magnetic field is anti-parallel to the real magnetic field in the bulk and parallel to the real magnetic field in the channels between the antidot and the walls of the constriction. Notation as in Fig.2. (b) Composite fermion Landau level structure and scattering configuration near e in Fig.5 (a) for a simple case. See text.

**Fig.6** Examples of conductance line shapes for an antidot in a constriction at zero temperature in (a) the fractional and (b) integer regime. In (a) $G_0$ is given by equations (8) and (26), while in (b) $G_{\text{int}} = e^2/h \; \Sigma_{ij} T_{ij}$ where $\Sigma_{ij} T_{ij}$ taken to be the same as $\Sigma_{ij} T_{ij}^*$ given by equation (26). The solid lines are for $T_e=0.9$ and $T_h=0.8$, while the dashed lines are for $T_e=0.05$ and $T_h=0.03$.

**Fig.7** Temperature dependence for some simple Aharonov-Bohm resonances associated with the edge states orbiting the antidot in Fig.5(a). $G_0$ is the zero temperature conductance, $G_T$ is the conductance calculated using the present composite fermion theory at a temperature $T=0.05\pi/(kc)$. $G_T^{\text{F}}$ is the prediction (12) of ordinary Fermi liquid phenomenology. (a) Simple reflection resonance with only scattering processes **e** and **h** active in Fig.5(a); $T_e=0.9$ and $T_h=0.8$. (b) Simple transmission resonance with only scattering processes **c** and **f** active in Fig.5(a); $T_c=0.9$ and $T_f=0.8$.